\documentstyle{mn}

\newcommand{\ie}{{\sl i.e. }}

\newcommand{\NaI}{$\rm Na~\sc I~$}
\newcommand{\CaII}{$\rm Ca~\sc II~$}
\newcommand{\chis}{$\chi^{2}~$}
\newcommand{\kms}{$\,$km$\,$s$^{-1}$}

\newcommand{\Msun}{\mbox{M$_{\odot}$}}

\title[The system parameters of the polars MR Ser and ST LMi]
{The system parameters of the polars MR Ser and ST LMi}
\author[T. Shahbaz and Janet H. Wood]{
T. Shahbaz$^{1}$ and Janet H. Wood$^{2}$ \\
$^{1}$University of Oxford, Department of Physics, Nuclear Physics
Laboratory, Keble Road, Oxford, OX1 3RH, UK \\
$^{2}$Keele University, Department of Physics, Keele, Staffordshire, 
ST5 5BG, UK}

\begin{document}

\maketitle

\begin{abstract}

\noindent
We obtain the \NaI $\lambda$8183,8195 absorption line radial
velocity curves for the polars ST LMi and MR Ser, from which we find the
semi-amplitudes to be $K_{abs}$=329$\pm$6 \kms\ and $K_{abs}$=289$\pm$9
\kms\ respectively. We find that for both systems the effects on the \NaI
absorption lines due to X-rays heating the inner face of the secondary
are negligible, and so the values obtained for $K_{abs}$ can be taken as
the true semi-amplitude of the secondary star. We then determine the
projected rotational velocities, $V_{rot}\sin i$, to be 104 $\pm$9 \kms\
and 66$\pm$13 \kms\ for ST LMi and MR Ser respectively which enables
their mass ratios to be calculated. For ST LMi and MR Ser we find the
mass ratio to be 0.22$\pm$0.04 and 0.10$\pm$0.05 respectively; values
which are significantly different only at the 94 percent level. Using the
value for the orbital inclination derived from polarimetric measurements,
we determine the mass, and the orbital and rotational velocities of the
secondary stars. These are significantly different at less than the 90
percent level. However, if the limb darkening is the same in both
objects, these quantities are significantly different at the 96 percent
level.

We present Doppler maps of the \NaI absorption and \CaII emission
in ST LMi and MR Ser. In both systems the \NaI absorption covers 
the secondary star. In ST LMi the Doppler image of the \CaII emission 
shows that it originates at the inner face of the secondary star. In MR Ser,
however, the emission lies close to but not on the secondary star. 

We show that ``spike'' in the orbital period distribution
of polars is a significant feature, although the discovery of only one
more system with a period outside the ``spike'' would decrease its
significance below a 99 percent confidence level. We conclude that, even
if the limb darkening coefficients for the secondary stars in ST LMi and
MR Ser are the same, we cannot rule out the two systems having identical
parameters. Therefore our observations are compatible with the theory
explaining the ``spike'' in the period distribution of the AM Hers.

\end{abstract}

\section{Introduction}

AM Her type systems consist of a magnetic white dwarf primary star
accreting material from a late type M dwarf secondary star. The white
dwarf primary star has a strong enough magnetic field (B$>$10 MG)
to lock it into synchronous rotation with the orbit. 
If one compares the orbital period distribution of the AM Hers with that of
the non-magnetic CVs, then one finds that the main features of the CV
distribution re-appear in the form of a minimum period $P_{orb}$=78 m,
and the period ``gap'' between $P_{orb}$=2-3 hrs (Wheatley 1995). 
However, there are three obvious features in the AM Her period
distribution (1) a concentration of systems near $P_{orb}$=114 m (the
``spike''), no such feature occurs in the distribution of weakly magnetic
CVs; (2) a long period cut-off, significantly shorter than that for CVs,
and (3) more systems in the period ``gap'' than for CVs.

Cataclysmic variables evolve from longer to shorter periods, ceasing mass
transfer as they pass through the period ``gap''. The period ``gap'' has been
modelled in terms of an interruption of orbital angular momentum loss at 3 hrs,
allowing the secondary star to relax into thermal equilibrium and detach from
its Roche lobe. Systems regain contact at 2 hrs after orbital angular momentum
loss through gravitational radiation. The ``spike'' in the polar distribution
has been explained as being the period at which most polars come out of the
period ``gap'' (Hameury, King \& Lasota 1990), However, it has been shown that
for the ``spike'' to be so narrow, for those systems in the ``spike'', the
masses of the secondary stars must ALL be 0.200 $\pm$ 0.002 $M_\odot$ AND the
masses of the white dwarfs must only differ by $\sim$ 0.05 $M_\odot$ (Ritter \&
Kolb 1992). Therefore not only do we expect the mass ratios for the systems in
the ``spike'' to be approximately the same, but the general system parameters
such as the orbital velocities should also be almost identical.
To test this theory the system parameters of polars and in particular
those in the spike must be determined more accurately than before. This
can be done by measuring the radial velocity of the secondary star,
$K_2$, and its rotational velocity, $V_{rot}\sin i$. In the red region of
the spectrum, where the M dwarf is expected to contribute significantly
to the observed flux, the spectra of polars are usually dominated by
cyclotron radiation, which originates in the post-shock flow near the
magnetic pole of the white dwarf, thus making observations of the red
component difficult. However, when they enter the low state and the
accretion deceases, or in parts of the orbit where the accreting region
is occulted it becomes much easier to observe the secondary star (see
Cropper 1990 for a review).


In this paper we report on the use of medium resolution, phase resolved
spectroscopy to study the orbital motion of the secondary stars in ST
LMi, MR Ser, AM Her and 1H 1752+081. We measure $V_{rot}\sin i$ and $K_2$
for MR Ser and ST LMi and deduce their system parameters. We also measure
$V_{rot}\sin i$ for AM Her and attempt to do so for 1H 1752+081. The
polars MR Ser, ST LMi and 1H 1752+081 lie in the ``spike''. We also
discuss the likelihood of the ``spike'' being a genuine feature and if
our observations of ST LMi and MR Ser are compatible with theoretical
explanations of the ``spike''.

\section{Observations and data reduction}

MR Ser, ST LMi, AM Her and 1H 1752$+$081 were observed with the ISIS
spectrograph attached to the 4.2-m William Herschel Telescope on the
nights of 1994 March 24/25 and 25/26. Here we only present the results
from the red arm. The EEV CCD was used with the R600R grating giving a
dispersion of 0.74 \AA\ pixel$^{-1}$. On the first night the wavelength
coverage was 7800--8586\AA\, but on the second night the central position was
moved to give 8030--8888\AA\ coverage. Our principal aim was to determine
radial velocities and to measure the rotational broadening of the
absorption line, so we used a narrow slit (1.0 arcsec) which resulted in
a spectral resolution of 1.9\AA\ (FWHM) $\equiv$70
\kms. Further observational details are contained in Table 1. The seeing
on the first night was steady at about 1 arcsec, and on the second night
it was typically between 0.5 and 1.2 arcsecs. The spectral type of the
secondary star in AM Her has been determined by many authors to be M5
(see Southwell et al. 1995 and references within). Mukai \& Charles
(1987) determined the spectral type of the secondary stars in ST LMi and
MR Ser to be M5.5. We therefore also observed Gl 406 (M6V) and Gl 473
(M5V) which we use for cross-correlation in the next section. We took
care to observe this star through the slit of the same width as used for
the two targets. The flux standard Feige 34 (Oke \& Gunn 1983) was
observed using wide and narrow slits on each night to correct for the
instrumental response. Cu-Ar arc exposures were taken every 30-40 min in
order to calibrate the wavelength scale. We binned the CCD pixels on
readout into groups of three pixels along the slit. The effect of this
was to reduce the readout noise and deadtime of the CCD.

Sensitivity variations were removed with a balance frame prepared from
tungsten-lamp flat-fields. Third order polynomial fits to the sky were
subtracted, and raw spectra were then extracted using the optimal
algorithm of Horne (1986), which weights the pixels along the spatial
direction in order to obtain the maximum signal-to-noise ratio. The arc
spectra were extracted with the same weights and at the same position as
the target, and fitted with third-order polynomials. The rms scatter of
the fits was $\sim$ 1/20th of a pixel. With four coefficients fitted to
about 20 lines in each case, the statistical uncertainly in the
wavelength scale is therefore of the order 1/45th of a pixel. This
corresponds to an uncertainty of 0.6 \kms.

Atmospheric features were removed from the red arm spectra by using a
wavelength calibrated water template star produced from the observed standard
star. The template star was then corrected to the zenith distance ($z$) of the
object by multiplying by a factor of (sec $z$)$^{0.6}$. A comparison of the
tabulated absolute flux values with a spline fit to the continuum of the
flux standard provided a correction for the large-scale instrumental
response. 

\section{Results}

\subsection{Radial velocities of ST LMi and MR Ser}

We obtained absorption line radial velocity curves using the \NaI
$\lambda$8183,8194 doublet and also emission line radial velocity curves
using the \CaII $\lambda$8498, 8542 lines. $K_{abs}$ and $K_{emi}$ are
defined as the semi-amplitude of the absorption and emission line radial
velocity curves respectively. The heliocentric radial velocities of MR
Ser and ST LMi are shown in Figures 1 and 2.

In order to measure the absorption line radial velocity curves we
cross-correlated the individual spectra with the template star spectrum
(Tonry \& Davis 1979), only using the wavelength range close to the \NaI
absorption lines.
Prior to cross-correlation, the spectra were
interpolated onto a logarithmic wavelength scale. Three spectra from ST
LMi and eighteen from MR Ser were too noisy to obtain a reliable
cross-correlation peak, and were not used for the fit. Gl 406 (M6V) was
used as the template spectrum. We find $K_{abs}$ to be
$329\pm6$~km~s$^{-1}$ and $289\pm9$~km~s$^{-1}$ for ST LMi and MR Ser,
respectively.

For each of MR Ser and ST LMi, to obtain the template spectrum for the
\CaII triplet emission lines we obtained a preliminary template by
averaging the spectra using $K_{abs}$. The summed spectrum showed some
double peaked structure implying that the velocity used to correct for
the Doppler motion was incorrect, and was not the velocity of the regions
producing the \CaII emission. We then performed the cross correlation to
obtain $K_{emi}$, only using the wavelength range which contained the 
\CaII $\lambda$8498, 8542 emission lines.
This value was then used to Doppler correct the
original spectra to give the final template spectrum for the \CaII
emission, in which the emission was narrow and single peaked. We find
$K_{emi}$ to be $213\pm4$~km~s$^{-1}$ and $173\pm1$~km~s$^{-1}$ for ST
LMi and MR Ser, respectively.

Because cataclysmic variables have very short periods, it is generally
believed that any initial non-circularity would have been quickly removed
by tidal forces between the red star and the white dwarf, and that the
present orbits are indeed circular. However, Davey and Smith (1992) argue
that the radial velocity curves may still be distorted from a pure sine
wave by geometrical distortion and heating of the secondary star by its
companion, causing the centre of light given by the strength of the \NaI
doublet to differ from the centre of mass. The effects of this can be
represented by allowing for a phase shift in the sine curve, or more
generally by introducing a fictitious eccentricity. They describe a
procedure for detecting any effects of heating on the radial velocity
curve. Firstly one must check for the significance of a fit with an
eccentric orbit. If the fit is not significantly better than a purely
sinusoidal fit, then the semi-amplitude of the curve $K_{abs}$ measured
from the absorption features represents a measure of the true
semi-amplitude of the radial velocity curve of the secondary star
$K_{2}$.

For both systems we found that the fit to the absorption line radial
velocity curve with an eccentric orbit was not significantly better than
that with a circular orbit. We obtained eccentricities of 0.027$\pm$0.017
and 0.075$\pm$0.046 for ST LMi and MR Ser, for which the significance of
the eccentricity parameter was less than 50 percent. Therefore the
measured value of $K_{abs}$ from the absorption lines can be taken to
represent the true value of $K_{2}$. After adding the heliocentric radial
velocity of the template star (13 \kms; Gliese 1969) we determined
$K_{2}$ using a circular fit of the form

 \begin{equation}
 V(\phi) = \gamma + K_{2} \sin (2\pi \phi)
 \end{equation}

\noindent
to the absorption and emission line heliocentric radial velocities (see
Table 2). Here $\gamma$ is the systemic velocity and $\phi$ is the
phase angle of the observations with $\phi$=0.0 at inferior conjunction
(i.e. when the red dwarf is in front of the white dwarf). The ephemerides
derived for ST LMi and MR Ser (see Table 2) agree well with those of
Mukai \& Charles (1987). Figures 3 and 4 show the Doppler corrected
average spectra of the two targets.

As another check to see if there is any effect due to heating, we fitted
the \NaI radial velocities measured only between phases $-$0.2 and +0.2,
when the secondary star is viewed from the back and its appearance should
be relatively unaffected by the effects of irradiation. We find that the
value for $K_{abs}$ does not change within the uncertainties. Our derived
values for $K_{2}$ from the sinusoidal fit to the radial velocity curve
are comparable with those derived by Mukai \& Charles (1987). However
there seems to be some discrepancy in the value for $\gamma$, which may
be explained by the fact that their are some inconsistencies between the
published radial velocity for most red dwarfs (Friend et al. 1990).
In order to check our $\gamma$ velocity estimates we performed the cross
correlation analysis using a fake template star, having the instrumental
resolution of the data, and the main absorption and emission lines
present at their rest wavelengths. This implies that the fake template
star has no $\gamma$ velocity associated with it. We found that our
derived parameters agreed well with those obtained earlier. Despite this,
it should be noted for the emission line case, that using the Doppler
averaged sum of the target star as a template star for the emission line
data may in some way bias the determination of the emission line $\gamma$
velocity. The result may depend on the phase chosen as the reference
phase for the average spectrum. Therefore, the only reliable estimate of
the $\gamma$ velocity from the emission lines is that obtained using the
fake template star.

\subsection{AM Her and 1H 1752$+$081}

Our data were less extensive for AM Her and 1H 1752+081. For AM Her we
cross-correlated the template star Gl 473 (M5V) with the target spectra
in order to determine the velocity shift. We then Doppler corrected and
summed the spectra. The mean spectrum is shown in Figure 5. For 1H
1752$+$081 the signal-to-noise ratio of the data was such that we were
unable to Doppler correct the spectra to the \NaI radial velocity. We
therefore merely show the summed spectrum in Figure 6. Only low
resolution blue spectra exist for this source (Ferrario et al. 1995).
Many emission lines can be seen in both objects such as $\rm He~\sc II~$
$\lambda$8246, \CaII $\lambda$8498, 8542, 8662 and the Paschen lines
$\lambda$8598, 8750. In AM Her the \NaI absorption lines are strong
whereas they are weak in 1H 1752+081. \NaI has not been observed in 1H
1752+081 previously.

\subsection{The rotational broadening and mass ratio}

In order to determine the rotational broadening, $V_{rot}\sin i$, we
followed the procedure described by Marsh, Robinson and Wood (1994). We
Doppler corrected and summed the spectra for our individual targets using
$K_{abs}$. We then fitted the excess light from the system i.e. that not
arising from the secondary star, plus a
constant times a rotationally broadened version of the template star, to
the object's spectrum. The template star was broadened by convolution
with the rotational profile of Gray (1976) assuming a linearized limb
darkening coefficient of $u=0.72$, taken from Al-Naimiy (1978). The
broadened templates and the excess light are shown for the individual
objects in Figures 3, 4 and 5. A high pass filter was then applied to the
difference spectrum to remove any large-scale spectral differences and
then the \chis was computed. Fitting a parabola to the distribution of
\chis versus broadening values, we obtain a minimum \chis for each target
and so the best value for the rotational broadening.

For ST LMi and MR Ser we obtained $V_{rot}\sin i$=105$\pm$7 and 66$\pm$12
\kms\ respectively, and for AM Her we obtained $V_{rot}\sin i$=70$\pm$11
\kms. The latter is consistent with that obtained by Southwell et al. (1995).
We were not able to determine $V_{rot}\sin i$ in 1H 1752+081. The
1-$\sigma$ uncertainties in $V_{rot}\sin i$ were derived by forcing the
minimum \chis to increase by 1 (Lampton, Margon \& Bowyer 1976).

The analysis above gives the uncertainties assuming the limb darkening is
known. However, in reality this is not the case. We therefore repeated
the above procedure using no and full limb darkening. For $u=0.0$ we find
$V_{rot}\sin i=98\pm7$, $60\pm12$ and $65\pm10$ \kms for ST LMi, MR Ser
and AM Her respectively. For $u=1.0$ we find $V_{rot}\sin i=110\pm7$,
$71\pm12$ and $72\pm12$ \kms for ST LMi, MR Ser and AM Her respectively.
We therefore adopt $V_{rot}\sin i$=104$\pm$9 \kms and $V_{rot}\sin
i$=66$\pm$13 \kms for ST LMi and MR Ser (see Figures 3 and 4). For AM Her
we find $V_{rot}\sin i$=68$\pm$12 \kms (see Figure 5).

Armed with $K_{2}$ and $V_{rot}\sin i$ we can now determine the mass
ratio $q(=M_{2}/M_{1})$ Assuming that the companion is tidally locked,
spherically symmetric and fills its Roche lobe, the rotational broadening
provides a measurement of the binary mass ratio through the result (see
e.g. Wade \& Horne 1988)

 \begin{equation}
 V_{rot} \sin i = 0.462 K_{2} q^{1/3}(1+q)^{2/3}.
 \end{equation}

\noindent
Substituting our values for $V_{rot}\sin i$ and $K_{2}$ in equation (2)
we find the mass ratios of ST LMi and MR Ser to be 0.22$\pm$0.04 and
0.10$\pm$0.05 respectively. (Since our results for AM Her are consistent
with those of Southwell et al. (1995) and they have already determined
the system parameters we do not do so here). The correction of
$V_{rot}\sin i$ and $q$ due to the Roche geometry only introduces an
error of a few percent (Marsh et al 1994; Southwell et al 1995). We
performed a $t$-test of the values obtained for the mass ratio for ST LMi
and MR Ser and find that the values for the mass ratios are only
significantly different at the 94 percent level (1.9$\sigma$).

However, if we assume that the two systems have the same limb darkening
coefficient, for example $u=0.72$, we then find the uncertainties in $q$
for ST LMi and MR Ser are reduced to 0.03 and 0.05 respectively. Using
these uncertainties, we then find that the mass ratios are significantly
different at the 96 percent level. We thus conclude that even if we
assume that the systems have similar limb darkening we cannot show at the
99 percent confidence level that the two systems have different mass
ratios.

\subsection{Doppler images}

The emission lines from cataclysmic variables are broadened by Doppler
shifting from the accretion disc (in weakly magnetic systems), component
stars and mass transfer. Marsh \& Horne (1988) have shown that it is
possible to derive the distribution of emission from observations of the
profile covering the binary orbit. Here we use the same technique but the
features we are interested in are narrow and arise from the companion
star or the accretion stream. The images are presented in velocity space
rather than position coordinates, since there is no unique transform
between velocity and position in accreting binaries.

Figures 7 and 8 show the Doppler maps of the \NaI absorption and \CaII
emission distribution for ST LMi and MR Ser respectively. We also place
on the maps the path of the gas stream and the shape of the secondary
star's Roche-lobe for the given values of $K_{2}$ and $q$. The \NaI maps
for both systems show absorption covering most of the secondary star, and
the \CaII emission arising from regions near the inner Lagrangian point
($L_{1}$). It should be noted that the resolution of the Doppler maps is
limited by the velocity dispersion of the data (see section 2).

The emission-line region is displaced towards the centre of mass of the
binary system, making the observed velocity amplitude smaller than the
true secondary star velocity. This ``$K_{2}$'' correction (Wade \& Horne
1988) can be expressed as

\begin{equation}
K_{2} = \frac{ K_{emi} }{ 1 - f(1+q)(R_{2}/a)}
\end{equation}

\noindent
where $R_{2}$ is the distance from the $L_{1}$ point to the centre of the
secondary star, and 0 $<$ f $<$ 1 expresses the distribution of line
emission over the surface of the secondary star. Placing the emission
entirely at the $L_{1}$ point i.e. $f$=1, for ST LMi using
$K_{emi}$=213$\pm$ 4 \kms\ we find $K_{2}$=313$\pm$13 \kms. This agrees
with the value for $K_{2}$ obtained from the absorption lines. For MR
Ser, the ``$K_{2}$'' correction gives a value for $K_{2}$ much smaller
that that obtained from the absorption line radial velocity curve. One
requires $q>$0.35 in order to obtain the observed $K_{2}$. These results
are consistent with the fact that the emission in the Doppler maps in ST
LMi lies on the secondary star near $L_{1}$, while in MR Ser it is close to
$L_{1}$ but not on the secondary star.

The accretion stream can be divided into two distinct parts. The first
part is free-falling in the orbital plane and the second is where the
stream grabs onto the magnetic field lines at the magnetospheric radius.
The phasing of the first narrow emission component would be offset from
that of the absorption lines arising from the secondary star. The
magnetic field in all AM Her type systems in which it has been measured
has been found to be similar, including ST LMi (Schmidt, Stockman \& Grandi
1983) and MR Ser (Wickramasinghe et al. 1991). For systems with periods
less than 2 hrs like ST LMi and MR Ser the magnetospheric radius is very
likely to be close to the $L_{1}$ point. Observations of ST LMi
show that the white dwarf magnetospheric radius is 0.79-0.93$R_{L1}$ and
in MR Ser $R_{mag}$ is expected to be very similar (Mukai 1988; where
$R_{L1}$ is the distance of the inner Lagrangian point from the white
dwarf). In MR Ser the emission we observe in velocity space is very close
to the $L_{1}$ point. It could arise from an extended emission region
around the $L_{1}$ point.
If it is due to emission from the region where the
free-falling gas stream joins onto the magnetic field lines of the
white dwarf, then one would expect it to lie on the gas stream in the
Doppler map. This is not the case.

\section{Discussion}

\subsection{The mass of the binary components}

Given our values of $K_{2}$ and $q$ combined with the orbital period, $P$, and
the binary inclination, $i$, we can also determine the masses of the compact
object and its companion using

\begin{equation}
\frac{K_{2}^{3} P }{2\pi G} = \frac{M_{1}\sin^{3}i}{(1+q)^{2}}.
\end{equation}

\noindent
The binary inclination for MR Ser and ST LMi can be obtained from
polarimetric observations. For ST LMi and MR Ser we use inclination
angles of $56^\circ \pm 4^{\circ}$ (Cropper 1988) and $45^\circ \pm
5^{\circ}$ (Brainerd \& Lamb 1985) respectively. We find that the masses
for the secondary stars in ST LMi and MR Ser are $M_{2}$=0.17$\pm$0.07
$M_{\odot}$ and $M_{2}$=0.07$\pm$0.08 $M_{\odot}$ respectively (see Table
2). Using the values for the inclination we can derive the orbital
velocity, $V_{orb}=K_2/\sin i$, and the rotational velocity, $V_{rot}$,
of the secondary star. For ST LMi and MR Ser we find $V_{orb}$ to be
397$\pm$20 \kms and 409$\pm$38 \kms, and $V_{rot}$ to be 125$\pm$12 \kms\
and 93$\pm$20 \kms respectively. If we assume that the two systems have
similar limb darkening, then the uncertainty in the $V_{rot}$
measurements reduces to 10 \kms and 19 \kms for ST LMi and MR Ser
respectively. Similarly, for $V_{orb}$ the uncertainties become 20 and 38
\kms and for $M_2$ 0.06 and 0.08 \Msun. The parameters are still not
significantly different at the 90 percent confidence level.

\subsection{The ``spike''}

The sample of known AM Hers has more than doubled over the last 5 years,
largely as a result of the ROSAT extreme ultraviolet and soft X-ray
surveys. To date there is a total of 43 systems (see Table 3). In Table 4
we list some details of how the AM Hers are distributed. We set the lower
and upper edges of the period ``gap'' to lie at 126 and 192 mins
respectively. The upper edge of the whole distribution is taken to be at
300 mins. The outlier RXJ0515+01 with an orbital period of 479 mins is
excluded from the analysis that follows because of its unusual nature.
Its low magnetic field strength is somewhat of a surprise, because for a
given white dwarf mass one would expect that the field strength required
to enforce synchronism should increase significantly as the orbital
period increases (Buckley \& Shafter 1995).

There are two questions we want to answer in order to explain the
observed characteristics of the period distribution (see Figure 9).

 \begin{itemize}
 \item What is the probability of the observed number of systems
lying in the period ``gap''? \\
 \item What is the probability of the  ``spike'' occurring by a 
  statistical fluke? 
 \end{itemize}

There are 42 AM Her systems in the period range used (see above), of
which 4 are in the period ``gap'' which lies between 126 and 192 mins.
Following the calculation of King 1994 (private communication) if the 
AM Hers are distributed uniformly in the range 78--300 mins, the
probability of finding at least 4 systems in the period ``gap'' is given by

 \begin{equation}
 P(\rm 4~in~``gap")=~^{N}C_{4}p^{N-4}(1-p)^{4}
 \end{equation}

\noindent
where $^NC_4$ is the number of ways of choosing 4 objects from a set of
$N$, $N$ is the total number of systems in the period range and $p$ is
the probability of an AM Her avoiding the ``gap''. Using $N$=42, we find
$p$=(1$-$66/222)=0.703, which implies P(4 in ``gap'')=1.3$\times 10^{-3}$,
\ie the observed number of systems in the gap is 
inconsistent with the distribution being uniform at the 99.9 percent level.

The pre-ROSAT data contained 12 members below the period ``gap'', of
which 6 were in a 2 min interval (the ``spike''). In the current sample
the number of ``spike'' members has risen to 9 at the cost of the
``spike'' broadening to 3 min. Suppose that there are $M$ AM Her systems
below the period gap, uniformly distributed over the period range of
$R$ mins, so that the probability of any one system being in a given
interval of 3 mins is 3/$R$. 
We want to know what is the probability
of the remainder $M$-1 being distributed in such a way that $r-1$ of
them (a total of $r$) will lie in the same 3 min interval, for any 
initial choice of interval. One has to add these results for each
of the initial choices, which means multiplying by $M$. Thus

 \begin{equation}
 P(``spike")=M ~^{M-1}C_{r-1}~p^{r-1}(1-p)^{M-r}.
 \end{equation}

\noindent
With $R$=38 mins corresponding to a range 78--116 mins (the most conservative),
$M$=25 and $r$=9 systems in the ``spike'', we find P(``spike'')=0.0068, \ie
the probability that the spike is genuine is 99.3 percent. Using the range
78--126 mins, $M$=27 and $r$=9 systems in the spike we find P(``spike'')=0.0030
with a probability of 99.7 percent that it is genuine. The old (pre ROSAT
launch) sample had 12 systems with 6 in a 1.2 min interval. Using data
before the ROSAT detections, we find the probability of the ``spike''
occurring randomly was 0.00014.

Although the significance of the ``spike'' has reduced since the ROSAT
mission, the ``spike'' is still a significant feature of the period
distribution of the AM Hers. However, it would take only one new system in the
period range 78--116 mins but outside the ``spike'' to raise the probability
that a 9 member ``spike'' could be a random event to more than 1 percent.

It should be noted that equation (6) is the probability that any
``spike'' with an interval of 3 mins occurs below the ``gap''. The
question we should next ask, is what is the probability of the ``spike''
with $r$ members occurring in a particular 3-min interval at 113--116
mins. This is given by

 \begin{equation}
 P(``spike"~at~113-116~mins)=~^MC_r p^{r}(1-p)^{M-r}
 \end{equation}

\noindent
Using $R$=38 mins (the most conservative), $M$=25 and $r$=9 systems in the
``spike'', we find P(``spike'' at 113--116 mins)=0.00007, \ie 99.99 percent.
Thus the ``spike'' in the period distribution occurring at 113--116 mins
is a significant feature.

If we compare the secondary star masses derived earlier for ST LMi and MR
Ser, with that predicted by the ``spike'' theory of $0.200 \pm 0.002
M_\odot$ (Ritter and Kolb 1992), we find that we cannot show with
certainty, i.e. at the 99 percent level, that the two systems have masses
different from this theoretical mass. If we accept the spike theory, then
the two systems should have almost identical system parameters, including
$q$, $M_2$, $V_{orb}$ and $V_{rot}$. The values we have obtained for all
the system parameters for ST LMi and MR Ser are such that we cannot rule
out that the two systems have the same parameters. Even if the limb
darkening coefficients are the same $q$ can only be shown to be different
at the 96 percent level. Our results are therefore compatible with the
theoretical explanation of the spike given by Ritter and Kolb (1992).
Given a value for the limb darkening the dominant source of uncertainty
in the deduced system parameters, other than $q$, is the orbital
inclination, $i$. To improve the sensitivity of this test of the spike
theory a more accurate determination of $i$ is required. Very high
resolution spectroscopy is needed to reduce the error in $V_{rot}\sin i$.
More systems in the spike should also be investigated.

\section{Conclusions}

We have obtained data of medium spectral and high time resolution
covering the orbital cycles of the polars ST LMi and MR Ser. Our spectra
allow us to determine the dwarf star's radial velocity semi-amplitude,
$K_{abs}$, using the \NaI $\lambda$8183,8195 absorption features. We find
that for both systems the effects of irradiation on the \NaI lines are
negligible, since there is no distortion of the radial velocity curves
from a pure sine wave. Therefore $K_{abs}$ can be taken as the true
semi-amplitude of the secondary star $K_{2}$. We find that for ST LMi
$K_{abs}=329\pm6$~km~s$^{-1}$ and for MR Ser
$K_{abs}=289\pm9$~km~s$^{-1}$. Doppler maps of the \NaI distribution for
the two systems are very similar, covering most of the secondary star.
However, in contrast, the \CaII maps are somewhat different. In ST LMi,
the map shows the emission arising from the secondary star at the inner
Lagrangian point, probably due to heating of the inner face of the
secondary star, whereas in MR Ser the emission is observed to arise from
regions close to but not on the secondary star. For ST LMi if we assume
all the emission arises from the $L_{1}$ point, then the semi-amplitude
of the emission line radial velocity curve gives a value for $K_{2}$
consistent with that obtained from the absorption line radial velocity
study.

The spectra of the secondary stars are significantly broadened by
rotation. We determine this projected rotational broadening which enables
us to determine the mass ratios of the systems. For ST LMi and MR Ser we
find $q$=0.22$\pm$0.04 and 0.10$\pm$0.05, respectively. Using the values
for the inclination derived from polarimetric measurements, we find
$M_{2}$=0.17$\pm$0.07 $M_{\odot}$, $V_{orb}=397\pm20$~km~s$^{-1}$ and
$V_{rot}=125\pm12$~km~s$^{-1}$ for ST LMi. For MR Ser we find
$M_{2}$=0.07$\pm$0.08 $M_{\odot}$, $V_{orb}=409\pm38$~km~s$^{-1}$ and
$V_{rot}=93\pm20$~km~s$^{-1}$. We conclude that we cannot rule out the
two systems having the same system parameters, therefore our observations
are consistent with the theoretical explanation of the spike.

Finally we show that at present the ``spike'' is a
significant feature of the period distribution of the AM Hers. However,
the discovery of only one more system with a period that does not lie in
the ``spike'' would decrease the significance of the spike below a 99
percent confidence level.

\section{Acknowledgements}

We are very grateful to our support astronomer Vik Dhillon for help above
and beyond the call of duty. We thank Tom Marsh for the use of his $\sc
pamella$, $\sc molly$ and $\sc doppler$ routines, and for his valuable
support and suggestions. We also thank Tom Marsh and Karen Southwell for
useful discussions. 
We thank the referee R.C. Smith for useful comments.
TS was supported by a PPARC postdoctoral fellowship,
and JHW by a PPARC Advanced fellowship. The WHT is operated on the island
of La Palma by the Royal Greenwich Observatory in the Spanish
Observatorio del Roque de los Muchachos of the Institutio de Astrofisica
de Canarias.
The data reduction wad carried out on the Oxford Starlink node.
Some of the figures were prepared using the $\sc ark$ software.

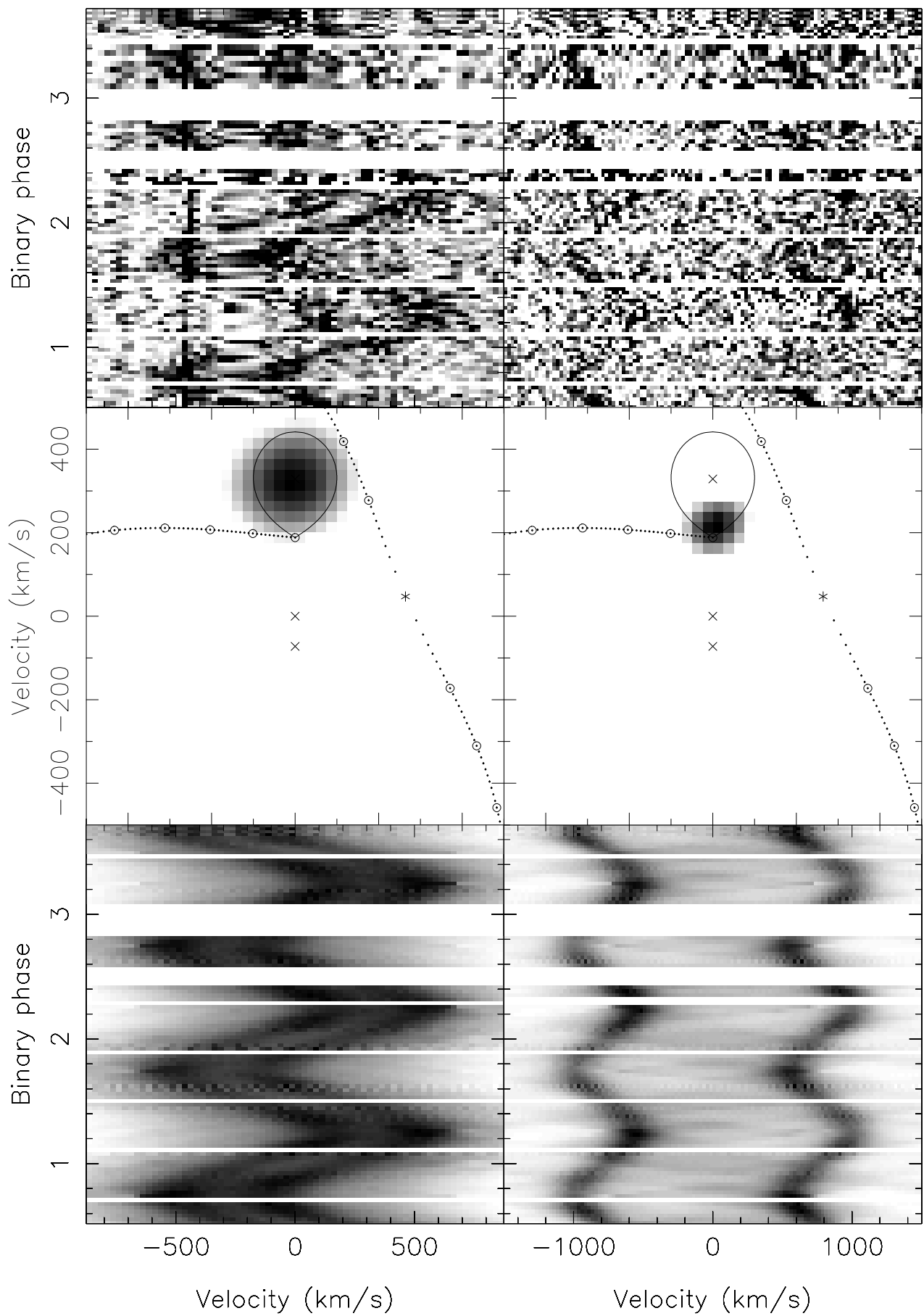
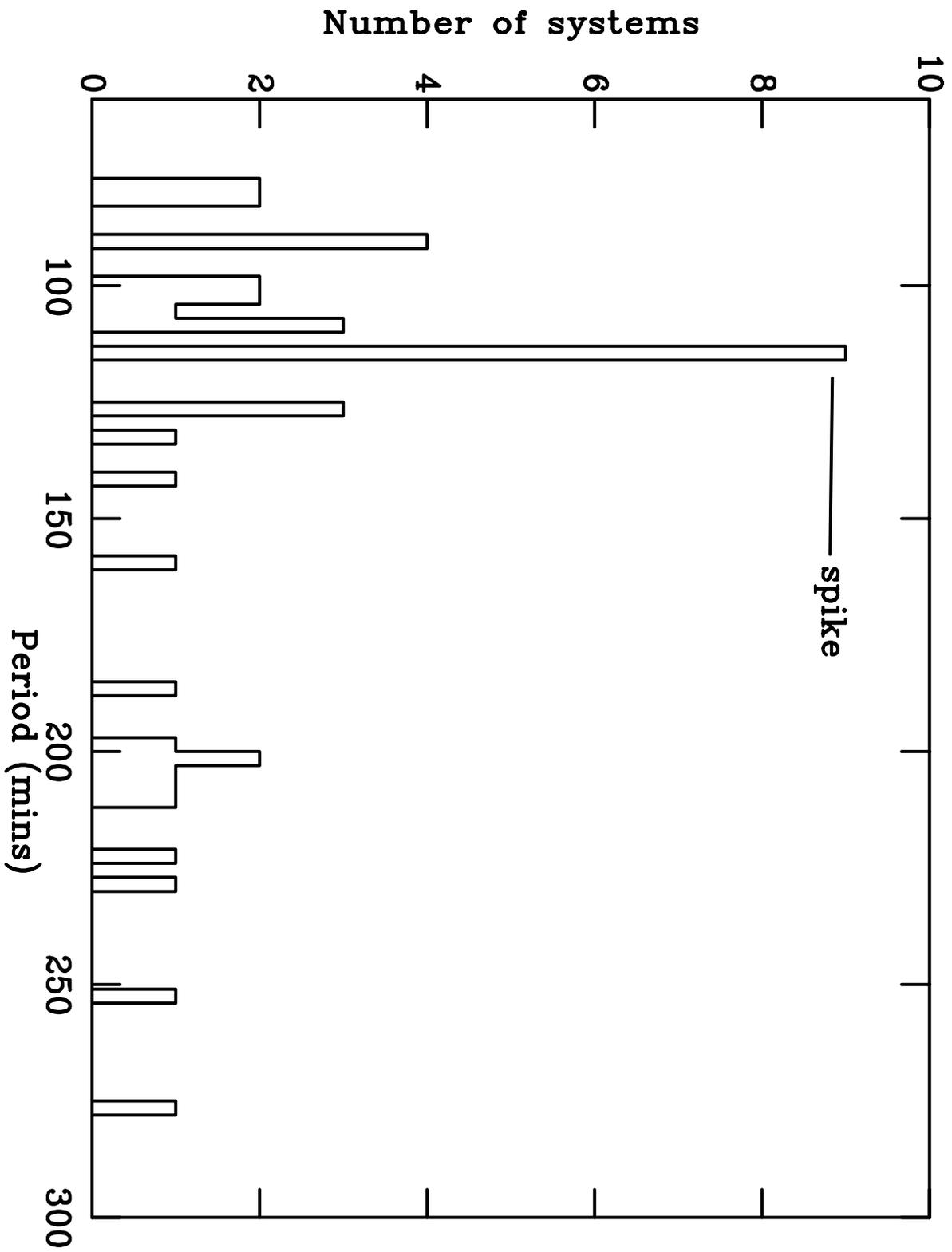
\end{document}